\documentclass{elsart}
\usepackage{epsfig}
\usepackage{psfrag}
\DeclareFontFamily{U}{msb}{}
\DeclareFontShape{U}{msb}{m}{n}{
<5><6><7><8><9> gen *msbm <10><10.95><12><14.4><17.28><20.74><24.88>msbm10}{}
\DeclareSymbolFont{AMSb}{U}{msb}{m}{n}
\DeclareMathSymbol{\AAA}{\mathbin}{AMSb}{'101}
\DeclareMathSymbol{\BBB}{\mathbin}{AMSb}{'102}
\DeclareMathSymbol{\CCC}{\mathbin}{AMSb}{'103}
\DeclareMathSymbol{\DDDD}{\mathbin}{AMSb}{'104}
\DeclareMathSymbol{\EEE}{\mathbin}{AMSb}{'105}
\DeclareMathSymbol{\FFF}{\mathbin}{AMSb}{'106}
\DeclareMathSymbol{\GGG}{\mathbin}{AMSb}{'107}
\DeclareMathSymbol{\HHH}{\mathbin}{AMSb}{'110}
\DeclareMathSymbol{\III}{\mathbin}{AMSb}{'111}
\DeclareMathSymbol{\JJJ}{\mathbin}{AMSb}{'112}
\DeclareMathSymbol{\KKK}{\mathbin}{AMSb}{'113}
\DeclareMathSymbol{\LLL}{\mathbin}{AMSb}{'114}
\DeclareMathSymbol{\MMM}{\mathbin}{AMSb}{'115}
\DeclareMathSymbol{\NNN}{\mathbin}{AMSb}{'116}
\DeclareMathSymbol{\OOO}{\mathbin}{AMSb}{'117}
\DeclareMathSymbol{\PPP}{\mathbin}{AMSb}{'120}
\DeclareMathSymbol{\QQQ}{\mathbin}{AMSb}{'121}
\DeclareMathSymbol{\RRR}{\mathbin}{AMSb}{'122}
\DeclareMathSymbol{\SSS}{\mathbin}{AMSb}{'123}
\DeclareMathSymbol{\TTTTT}{\mathbin}{AMSb}{'124}
\DeclareMathSymbol{\UUU}{\mathbin}{AMSb}{'125}
\DeclareMathSymbol{\VVVV}{\mathbin}{AMSb}{'126}
\DeclareMathSymbol{\WWW}{\mathbin}{AMSb}{'127}
\DeclareMathSymbol{\XXXX}{\mathbin}{AMSb}{'130}
\DeclareMathSymbol{\YYY}{\mathbin}{AMSb}{'121}
\DeclareMathSymbol{\ZZZZ}{\mathbin}{AMSb}{'132}
\newcommand{\XX}{{\bf r}}

\newcommand{\EX}{{\sf EX}}

\newcommand{\GF}{{\sf GF}}

\begin{document}
\begin{frontmatter}
\title{Microstructure analysis of reconstructed porous media}
\author{B. Biswal$^{1,2}$ and R. Hilfer$^1$}
\address{$^1$ICA-1, Universit{\"a}t Stuttgart, 
Pfaffenwaldring 27, 70569 Stuttgart\\
$^2$Department of Physics \& Electronics, Sri Venkateswara College,\\
University of Delhi, New Delhi - 110 021}

\begin{abstract}We compare the quantitative microstructural properties
  of Berea Sandstone with stochastic reconstructions of the same
  sandstone. The comparison is based on local porosity theory. The
  reconstructions employ Fourier space filtering of Gaussian random
  fields and match the average porosity and two-point correlation
  function of the experimental model. Connectivity properties of the 
  stochastic models differ significantly from the experimental model. 
  Reconstruction models with different levels of coarse graining 
  also show different average local connectivity.
\end{abstract}
\end{frontmatter}

Recently a number of stochastic models have been 
proposed for reconstruction of the microstructure of porous 
media(see \cite{adl92,YT98} and references therein). 
To assess the quality of the reconstruction, 
it is neccessary to have quantitative methods of comparison for such
microstructures. General geometric characterization methods 
normally include porosities, specific surface areas and 
correlation functions \cite{hilf96}. 
Here we follow a more general quantitative 
characterization for stochastic microstructures which is based on 
local porosity theory (LPT) \cite{hil92a,hilf96}. 
Our analysis allows to distinguish quantitatively between three
different microstructures all of which have identical porosities
and correlation functions.
The three microstructures are an
experimental sample of Berea Sandstone obtained by computerized
microtomography and two stochastic models of the same sandstone
obtained through the Gaussian filtering method \cite{adl92}. 

Consider a three-dimensional sample 
$\SSS=\PPP\cup\MMM$ (with $\PPP\cap\MMM=\emptyset$)
where $\PPP$ is the pore space, $\MMM$ is the 
rock or mineral matrix. $\emptyset$ denotes the empty set.
The porosity $\phi(\SSS)$ of such a two component porous medium
is defined as the ratio $\phi(\SSS) = V(\PPP)/V(\SSS)$
where $V(\PPP)$ denotes the volume of the pore space,
and $V(\SSS)$ is the total sample volume. For the sample 
data analysed here the set $\SSS$ is a cube
with sidelength $M$ in units of the 
lattice constant $a$ of a simple cubic
lattice. 
Let $\KKK(\XX,L)$ denote a cube of sidelength $L$
centered at the lattice vector $\XX$.
The set $\KKK(\XX,L)$ defines a measurement cell
inside of which local geometric properties such
as porosity and pore space connectivity are
measured \cite{hil92a,hilf96,bmh98}.
The {\it local porosity} in this measurement cell $\KKK(\XX,L)$ 
is defined as
$\phi(\XX,L)=[V(\PPP\cap\KKK(\XX,L))]/[V(\KKK(\XX,L))]$.
The {\it local porosity distribution} $\mu(\phi,L)$ is defined as
$\mu(\phi,L) = \frac{1}{m}\sum_\XX\delta(\phi-\phi(\XX,L))$, where $m$ 
is the number of placements of the measurement cell $\KKK(\XX,L)$.
For better statistics the results presented here are obtained 
by placing $\KKK(\XX,L)$ on all lattice sites $\XX$ which are 
at least a distance $L/2$ from the boundary of $\SSS$.
The {\it local percolation probabilities} characterize the connectivity
of measurement cells of a given local porosity.
Let $\Lambda_\alpha(\XX,L)$ equal $1$ if $\KKK(\XX,L)$ percolates in
``$\alpha$'' direction and $0$ otherwise, be an indicator for percolation. 
A cell $\KKK(\XX,L)$ is called ``percolating in the $x$-direction'' 
if there exists a path inside the set
$\PPP\cap\KKK(\XX,L)$ connecting those two faces
of $\SSS$ that are vertical to the $x$-axis.
Similarly for the other directions.
$\Lambda_3=1$ indicates that the cell can be traversed along 
all 3 directions, while $\Lambda_c=1$ indicates that there exists
at least one direction along which the block is percolating. The 
local percolation probability in the ``$\alpha$''-direction
is defined through
$\lambda_\alpha(\phi,L) = [{
\sum_\XX \Lambda_\alpha(\XX,L)\delta_{\phi\phi(\XX,L)}}]/
[{\sum_\XX\delta_{\phi\phi(\XX,L)}}]$
and gives the fraction of measurement
cells of size $L$ having porosity $\phi$ that are percolating in the ``$\alpha$''-direction.
The {\it total fraction of percolating cells} which percolate along the
``$\alpha$''-direction is given by 
$p_\alpha(L)=\int_0^1\mu(\phi,L)\lambda_\alpha(\phi,L)\;d\phi$.
 
The Gaussian field ({\sf GF}) reconstruction model \cite{adl92} 
generates a random pore space configuration with inputs from 
a given experimental sample. Given the reference correlation 
function $G_\EX(\XX)$ and porosity $\phi(\SSS_\EX)$ 
of the experimental sample, the three main steps of 
constructing the sample $\SSS_\GF$ with correlation function
$G_\GF(\XX)=G_\EX(\XX)$ and porosity $\phi(\SSS_\GF)=\phi(\SSS_\EX)$ are 
as follows:
\begin{enumerate}
\item
A standard Gaussian field $X(\XX)$ is generated which consists of 
statistically independent Gaussian random variables $X\in\RRR$ at each 
lattice point $\XX$.
\item
\label{l2}
The field $X(\XX)$ is first passed through a linear filter
which produces a correlated Gausssian field $Y(\XX)$ with
zero mean and unit variance.
\item
The correlated field $Y(\XX)$ is then passed through a nonlinear
discretization filter which produces the reconstructed sample $\SSS_\GF$.
\end{enumerate}

For the process described in step \ref{l2}, 
we have followed an alternate and computationally more efficient 
method proposed in Ref. \cite{adl92} that uses Fourier Transforms. 
An effective reconstruction 
requires a large separation $\xi_\EX\ll M$ where $M$ is the
sidelength(in pixels) of the sample and $\xi_\EX$ is the
correlation length of the experimental reference, defined as the
length such that $G_\EX(r)\approx 0$ for $r>\xi_\EX$.
Violation of this condition leads to inaccuracy in the implementation
of step \ref{l2} of the reconstruction, which in turn leads to a discrepancy 
at small $r$ between $G_\GF(r)$ and $G_\EX(r)$. This problem can be 
overcome by choosing large $M$. However, in $d=3$ very large $M$ 
also demands prohibitively large memory. Apart from this, the
reconstruction also depends crucially on two other parameters, a
length $M_c$ up to which the experimental correlation is 
incorporated into the reconstructed sample, and $n$, an interval at
which the $G_\EX(r)$ is sampled. For better reconstruction
$G_\EX(M_c)$ needs to be negligibly small.
Different values of $n$ correspond to a change of length scale. 
The model BR1 is constructed with $n=1$ and BR2 
with $n=2$. 
Although both have the same sidelength, the effective 
size of BR2 is twice that of BR1 because of this coarse graining 
procedure.

In Fig.1 the averaged correlation functions 
$G(r)=(G(r,0,0)+G(0,r,0)+G(0,0,r))/3$ for the three samples are 
plotted.
The experimental sample is a Berea sandstone of porosity 
$\phi(\SSS_{EX})=0.1775$ \cite{bmh98}. The model BR1, 
with $M=128$, $n=1$ and 
$M_c=32$, shows descrepancies for small values of $r$. 
However, for BR2, with $M=128$, $n=2$ and $M_c=32$, the 
correlation function matches more closely to that of EX. 
The resolution $a$ of the experimental sample EX is $10\mu m$. 
Hence the actual size of EX and BR1 is $1280\mu m$, whereas
that of BR2 is $2560\mu m$. The porosities 
match quite well for all the samples ($\phi(\SSS_{BR1})=0.1783$ 
and $\phi(\SSS_{BR2})=0.1776$). 

The reconstructed models $BR1$ and $BR2$ are isotropic and
globally connected, i.e., the pore spaces percolate in all the 
three directions.
The local porosity analysis results are plotted in Fig. 2 for the
experimental sample EX, in Fig. 3 for the stochastic model BR1, and in
Fig. 4 for BR2.  
Comparison of $\mu(\phi,L)$ indicates that the stochastic models 
have nearly the same level of homogeneity with that of the 
experimental sample.
The main differences are found in
$\lambda_\alpha(\phi,L)$ of the stochastic models. They
differ significantly from that of EX, and they also vary widely among
themselves. The reconstructed models have lower average connectivity
of pore spaces. We observe $[\lambda_\alpha(\phi,L)]_{EX}~>~
[\lambda_\alpha(\phi,L)]_{BR1}~>~[\lambda_\alpha(\phi,L)]_{BR2}$.
These differences appear even more clearly in the plot of 
$p_\alpha (L)$(inset of Fig 2, Fig 3 and Fig 4).  
In the experimental model(Fig 2) nearly all the measurement 
cells of dimension larger than $400\mu m$ are percolating(globally
connected pore space) in all directions.   
Comparison of 
$p_3 (L)$ of the three models(Fig. 5) shows the drastic loss of
average connectivity of the reconstructed models. 
Fig. 5 shows that the {\sf GF} reconstruction BR1 with $n=1$ has a lower
connectivity than EX. In $BR1$ nearly
60\% of the measurement cells of size $400\mu m$ 
percolate in all directions. Coarse graining \cite{adl92} to $n=2$ further
reduces the connectivity of pore space. In $BR2$ less than
30\% of the measurement cells of size $400\mu m$ 
percolate in all directions. These results indicate 
that gaussian filtering  reconstruction methods retain a similar degree
of isotropy and homogeneity as the original sandstone but 
tend not to reproduce connectivity properties that are important for
transport.

\newpage
\begin{figure}
\psfrag{r}{\Large $r~(\mu m)$}
\psfrag{gr}{\Large $G(r)$}
\begin{center}
\epsfig{figure=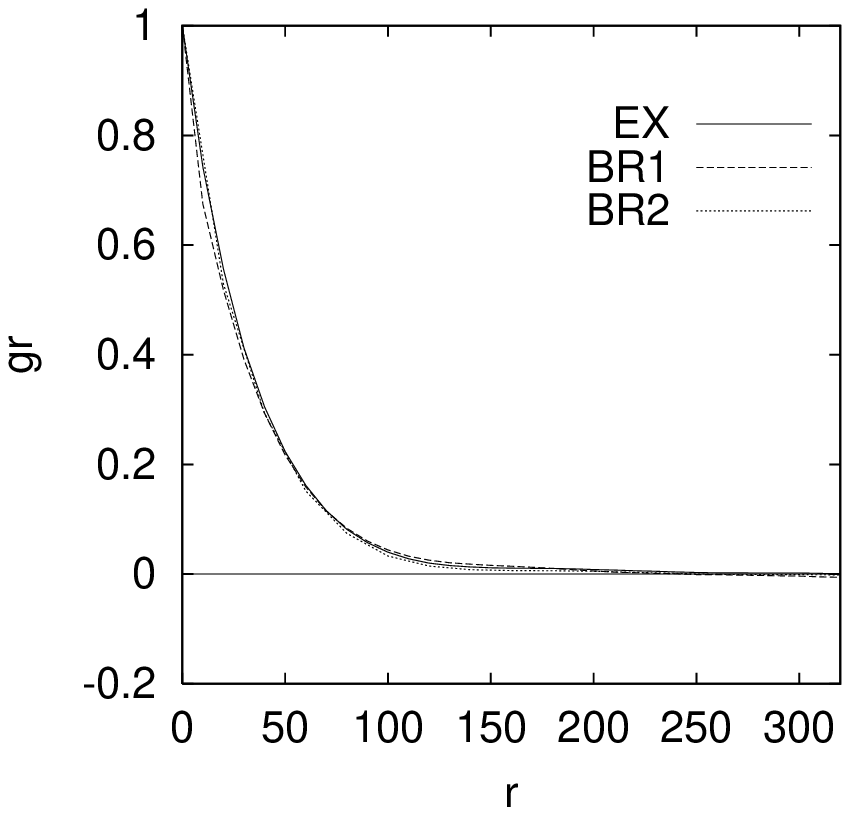,angle=0}
\end{center}
\caption{Averaged directional correlation functions 
of all three models}
\label{fig1}
\end{figure}

\newpage
\psfrag{xlabel}{\Large $\phi$}
\psfrag{ylabel}{\Large $\lambda_\alpha(\phi;L)$}
\psfrag{y2label}{\Large $\mu(\phi;L)$}
\psfrag{p(L)}{ $p_\alpha(L)$}
\psfrag{L}{ $L$}
\psfrag{pl3}{ $p_3(L)$}
\psfrag{x1}{0}
\psfrag{x2}{ }
\psfrag{x3}{200}
\psfrag{x4}{ }
\psfrag{x5}{400}
\psfrag{y1}{0.0}
\psfrag{y2}{.25}
\psfrag{y3}{.50}
\psfrag{y4}{.75}
\psfrag{y5}{1.0}
\begin{figure}
\begin{center}
\epsfig{figure=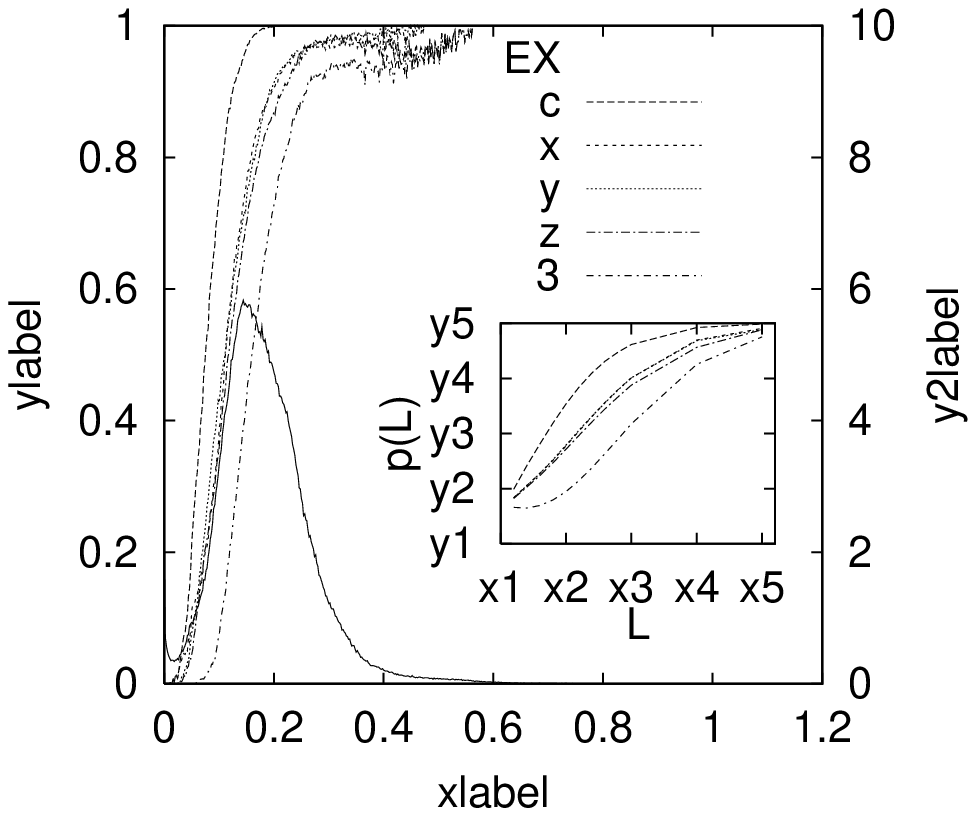,angle=0}
\end{center}
\caption{
$\lambda_\alpha(\phi,L)$ (broken curves, left axis) and 
$\mu(\phi,L)$(solid curve, right axis) at $L=200\mu
m$ for the model EX. The inset shows the function $p_\alpha (L)$,
$\alpha$ is given in the legend. }
\end{figure}

\newpage
\begin{figure}
\begin{center}
\epsfig{figure=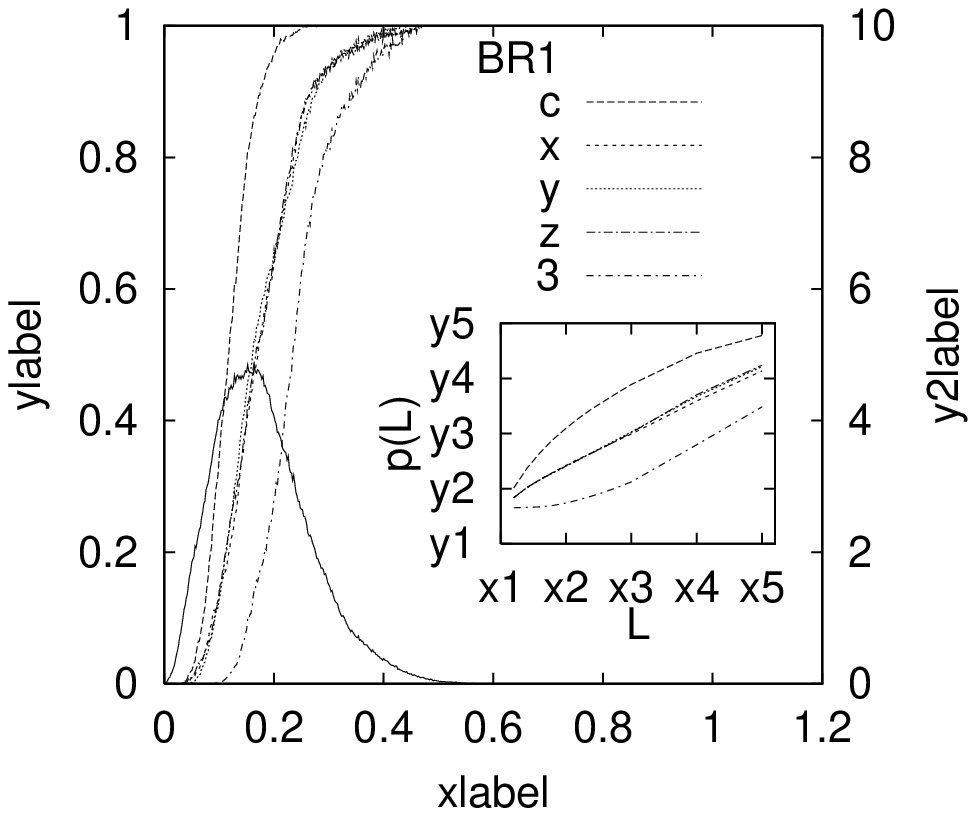,angle=0}
\end{center}
\caption{
$\lambda_\alpha(\phi,L)$ (broken curves, left axis) and 
$\mu(\phi,L)$(solid curve, right axis) at $L=200\mu
m$ for the model BR1. The inset shows the functions $p_\alpha (L)$,
$\alpha$ is given in the legend.}
\end{figure} 

\newpage
\begin{figure}
\psfrag{x1}{0}
\psfrag{x3}{400}
\psfrag{x5}{800}
\begin{center}
\epsfig{figure=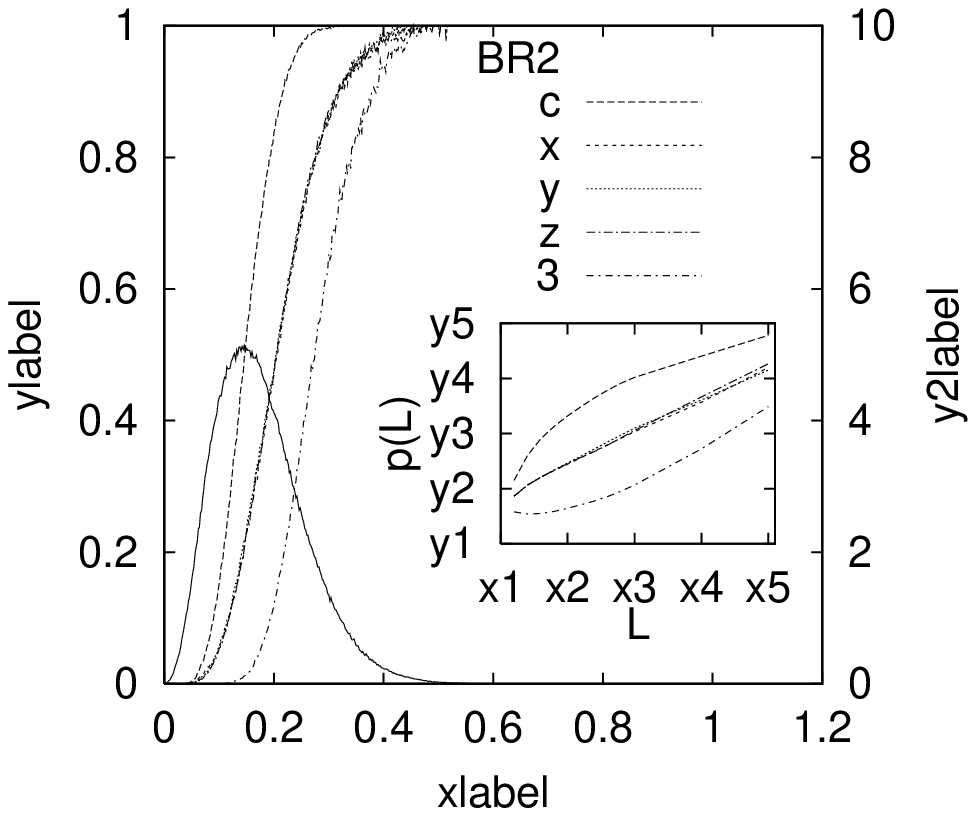,angle=0}
\end{center}
\caption{
$\lambda_\alpha(\phi,L)$ (broken curves, left axis) and 
$\mu(\phi,L)$(solid curve, right axis) at $L=200\mu
m$ for the model BR2. The inset shows the functions $p_\alpha ((L)$,
$\alpha$ is given in the legend.}
\end{figure} 

\newpage 
\psfrag{50}{}
\psfrag{150}{}
\psfrag{250}{}
\psfrag{350}{}
\psfrag{450}{}
\begin{figure}
\begin{center}
\epsfig{figure=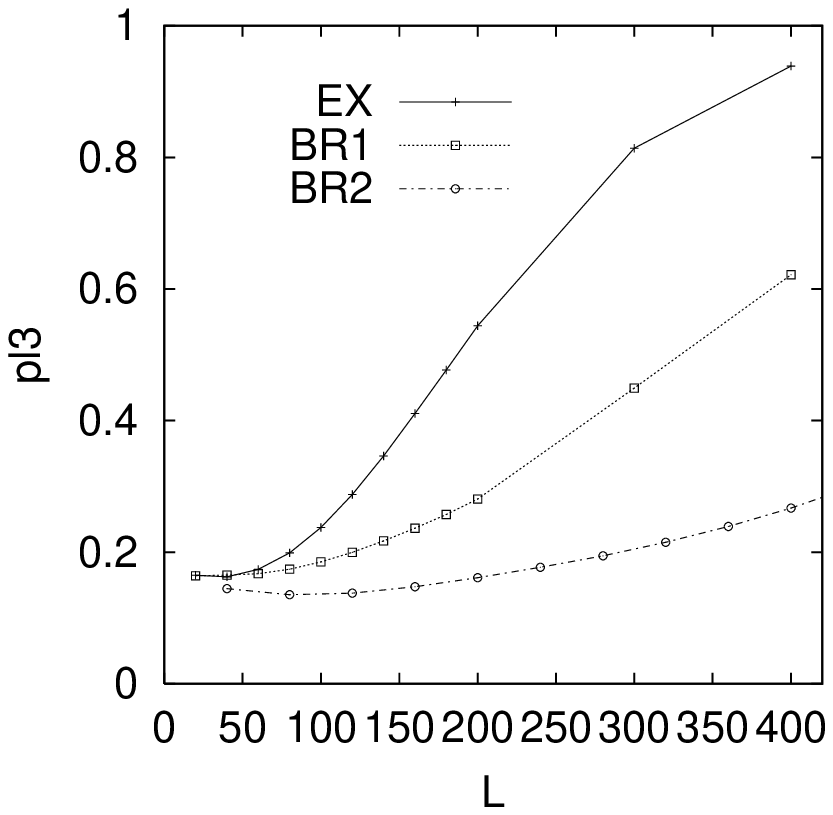,angle=0}
\end{center}
\caption{$p_3(L)$ for the three samples}
\end{figure} 


\begin{thebibliography}{9}
\bibitem{adl92}
P. Adler, 
{\em Porous Media\/}.
(Boston: Butterworth-Heinemann, 1992).
\bibitem{YT98}
C. Yeong and S. Torquato, 
Reconstructing random media,
{\em Phys. Rev. E\/} {\bf 57} (1998) 495--506.
\bibitem{hil92a}
R. Hilfer, ``Local Porosity Theory for Flow in Porous Media'', 
{\em Phys.Rev. B\/} {\bf 45} (1992) 7115--7121.
\bibitem{hilf96}
R. Hilfer,
Transport and relaxation phenomena in porous media,
{\em Advances in Chemical Physics\/} {\bf XCII} (1996) 299--424.
\bibitem{bmh98}
B. Biswal, C. Manwart and R Hilfer,
Three-dimensional local porosity analysis of porous media, 
{\em Physica A\/} {\bf 255} (1998) 221-241
\end{thebibliography}
\end{document}